\documentclass[useAMS,usenatbib]{mn2e}
\usepackage{epsfig}
\usepackage{graphicx}
\usepackage{subfigure}

\title[e-VLBI observations of Circinus X-1 on AU scales]{e-VLBI observations of Circinus X-1: monitoring of the quiescent and flaring radio emission on AU scales}
\author[A. Moin et al.]{A. Moin$^{1,2}$\thanks{E-mail:
aquib.moin@postgrad.curtin.edu.au (CIRA/ICRAR)}, C. Reynolds$^{1}$, J.~C.~A. Miller-Jones$^{1}$, S.~J. Tingay$^{1}$, C.~J. Phillips$^{2}$, \and A.~K. Tzioumis$^{2}$, G. D. Nicolson$^{3}$, R.~P. Fender$^{4}$ \\ $^{1}$International Centre for Radio Astronomy Research, Curtin University, Bentley 6102, WA, Australia \\ $^{2}$CSIRO Astronomy and Space Science, PO Box 76, Epping, NSW, Australia \\ $^{3}$ Hartebeesthoek Radio Astronomy Observatory, PO Box 443, Krugersdorp 1740, South Africa \\ $^{4}$School of Physics and Astronomy, University of Southampton, High Field SO17 IBJ, England} 

\begin{document}

\date{Accepted 2011 xxxx. Received 2010 xxxx; in original form 2010 December}

\pagerange{\pageref{firstpage}--\pageref{lastpage}} \pubyear{2011}

\maketitle

\label{firstpage}

\begin{abstract}
A recent detection of the peculiar neutron star X-ray binary Circinus X-1 with electronic very long baseline interferometry (e-VLBI) prompted the suggestion that compact, non-variable radio emission persists through the entire 16.6-day orbit of the binary system.  We present the results of a high angular resolution monitoring campaign conducted with the Australian Long Baseline Array in real-time e-VLBI mode. e-VLBI observations of Circinus X-1 were made on alternate days over a period of 20 days covering the full binary orbit. A compact radio source associated with Circinus X-1 was clearly detected at orbital phases following periastron passage but no compact radio emission was detected at any other orbital phase, ruling out the presence of a persistent, compact emitting region at our sensitivity levels.  The jet was not resolved at any epoch of our 1.4-GHz monitoring campaign, suggesting that the ultrarelativistic flow previously inferred to exist in this source is likely to be dark.  We discuss these findings within the context of previous radio monitoring of Circinus X-1.

\end{abstract}

\begin{keywords}
Circinus X-1,  X-ray binary, Neutron star, VLBI, jets, radio continuum, accretion disk
\end{keywords}

\section{Introduction}

Circinus X-1 (Cir X-1) is a Galactic X-ray binary (XRB) system consisting of a confirmed low magnetic field neutron star \citep{Lin10} accreting matter from a less evolved companion believed to be a supergiant of spectral type B5--A0 \citep{Jon07}.  At a distance of 3.8--10.5\,kpc \citep{Jon04,Iar05}, the binary system has an eccentric orbit with an orbital period of 16.6\,d \citep{Kal76}, although the exact eccentricity is not well constrained, with estimates ranging from $e\sim0.45$ \citep{Jon07} to  $e\sim0.8$ \citep{Mur80}. The enhanced mass accretion rate onto the neutron star close to periastron is believed to explain the X-ray, infrared and radio flares seen following orbital phase 0 \citep{Mur80,Hay80}.  During these flares, Cir X-1 can become the brightest confirmed neutron star system at radio wavelengths, with flux densities $>1$\,Jy.

Depending on their mass accretion rate, accreting low-magnetic field neutron stars show different sets of X-ray spectral and timing characteristics, according to which they are classified as either atoll sources or Z-sources \citep{Hom10}.  Cir X-1 can show the properties of both Z sources \citep{Shi97} and atoll sources \citep{Oos95}, and prior to periastron passage shows behaviour that is characteristic of neither class \citep{Sol09}.  The source thus defies simple classification.

Cir X-1 is surrounded by a synchrotron nebula, which is believed to have been inflated by jets launched from the inner regions of the accretion flow \citep{Tud06}.  The jets have been resolved on arcsecond scales in both the radio \citep{Ste93,Tud06} and X-ray \citep{Hei07,Sol09a} bands. The time delay between the successive brightening of different radio components was interpreted as evidence for the relativistic nature of the outflow \citep[$\Gamma > 15.0$ for a distance of 6.5\,kpc;][]{Fen04,Tud08}.

The intensity of the radio flares close to periastron has varied significantly over time.  In the period 1975--1985, peak intensities in excess of 2\,Jy were observed at radio wavelengths \citep[e.g.][]{Hay78}.  After 1985, the flares became significantly weaker, peaking at $<50$\,mJy in the period 1996--2006 \citep{Tud08}.  More recently, the source has shown episodes of brighter flaring events \citep{Nic07,Cal10}, following which electronic very long baseline interferometry (e-VLBI) observations with the Long Baseline Array (LBA) detected compact radio emission from Cir X-1 on milliarcsecond scales \citep{Phi07} for the first time since the high-activity phase of 1975--1985 \citep{Pre83}.

The appearance of the compact radio source associated with Cir X-1 in 2007, as revealed with e-VLBI, was suggested to be the quiescent, non-variable emission from the binary system that is persistent during at least part of the 16.6 day orbit, and was not associated with a high flaring state of the system.  This conclusion was based on a comparison of the data from \citet{Phi07} and \citet{Pre83}, in particular the estimated sizes of the radio emitting structures.

In order to determine the evolution of the radio emission as a function of the binary orbital phase, \citet{Phi07} suggested an extended campaign of sensitive e-VLBI observations could be used to distinguish constant quiescent radio emission from the flaring radio emission seen near periastron.  This paper reports on the results of such an extended campaign.

\section{Observations \& Results}

An e-VLBI observing campaign was conducted with multiple observing sessions in 2009 February/March, July and December, to follow the evolution of the compact radio emission associated with Cir X-1 around its 16.6 day orbit.   We present here the observed flux densities obtained in 2009, to improve our sampling of the compact radio emission as a function of orbital phase. Table~\ref{tab:evlbi_obs} summarizes the data obtained during the multiple observing sessions.  The All Sky Monitor (ASM) on board the {\it Rossi X-ray Timing Explorer} ({\it RXTE}) satellite indicated that the one-day averaged 2-10\,keV X-ray count rate over the period of our observations was always $<5$\,c\,s$^{-1}$, with an average below 1\,c\,s$^{-1}$.  This is over an order of magnitude lower than observed in the period 1996--2003 \citep[see fig.~1 of][]{Tud08}, when the flaring activity was already significantly weaker than observed prior to 1985.  The radio flares at periastron were consequently significantly weaker during our observing campaign than during the only previous orbital phase resolved VLBI monitoring campaign of \citet{Pre83}.

\begin{table*}
\centering
\begin{minipage}{140mm}
\caption{Newly-reported e-VLBI observations of Cir X-1.  In the case of non-detections, the flux densities reported are the $3\sigma$ upper limits.}
\begin{tabular}{@{}llrrrrlrlrl@{}}
\hline
Date     &   Time    & Orbital & Beam size & Image rms noise & Flux density  & Frequency\\
                   & UT       &  Phase & (mas $\times$ mas)  & (mJy\,beam$^{-1}$)                 & (mJy)                   & (GHz) &\\ \hline 
 26/02/2009 & 17:30-21:30 & 0.768-0.779 &   $141\times75$ &  0.14 & $<0.42$  & 1.4  \\
 01/03/2009 & 13:30-20:00 & 0.940-0.956 &   $163\times77$ & 0.20 & $<0.60$  & 1.4  \\ 
 15/07/2009 & 04:00-07:00 & 0.150-0.157 &   $727\times50$ & 0.56 & $5.6\pm0.6$    & 1.7        \\ \hline
 30/11/2009 & 19:00-22:00 & 0.541-0.548 &   $403\times63$  & 0.10 & $<0.30$  & 1.4  \\
 02/12/2009 & 19:00-22:00 & 0.661-0.669 &   $753\times61$ & 0.12 & $<0.36$  & 1.4  \\
 04/12/2009 & 19:00-22:00 & 0.782-0.790 &   $452\times62$ & 0.25 & $<0.75$  & 1.4  \\
 06/12/2009 & 19:00-22:00 & 0.903-0.911 & $1060\times61$ & 0.13 & $<0.39$  & 1.4\\
 09/12/2009 & 19:00-22:00 & 0.086-0.094 &   $1360\times62$ & 0.54 & $3.7\pm0.5$  & 1.4 \\
 11/12/2009 & 19:00-22:00 & 0.207-0.215 & $164\times20$ & 0.31 & $2.4\pm0.3$  & 1.4\\
 16/12/2009 & 19:00-22:00 & 0.509-0.517 &   $306\times20$ & 0.37 & $<1.11$ & 1.4\\ 
\hline
\label{tab:evlbi_obs}
\end{tabular}
\end{minipage}
\end{table*}

The telescopes used in the e-VLBI array were the Parkes radio telescope (64m), Australia Telescope Compact Array (ATCA) in Narrabri (5 x 22m) and the Mopra radio telescope (22m), all operated by Australia's Commonwealth Scientific and Industrial Research Organisation (CSIRO) Astronomy and Space Science division. The Hobart 26-m telescope, operated by The University of Tasmania, was used for two sessions conducted on 2009 December 9 and 16. The three CSIRO telescopes were connected via a 1 Gbps network and data were transported from the telescopes to a Beowulf cluster at the ATCA. Data were streamed from the ATNF telescopes at a sustained rate of 512 Mbps from the telescopes to the processing cluster and the data from Hobart were streamed at 128 Mbps. The cluster at the ATCA ran the DiFX software correlator \citep{Del07}, which was used to correlate the data streams coming from the telescopes in real time. The observing sessions in 2009 February, March and December were conducted at a frequency of 1.4\,GHz, while the frequency of observations in 2009 July was 1.7\,GHz. The total bandwidth for each of the observing sessions was 64 MHz in each of two orthogonal circular polarizations recorded as $4\times16$\,MHz bands.

\begin{figure}
\includegraphics[width=3in,angle=0]{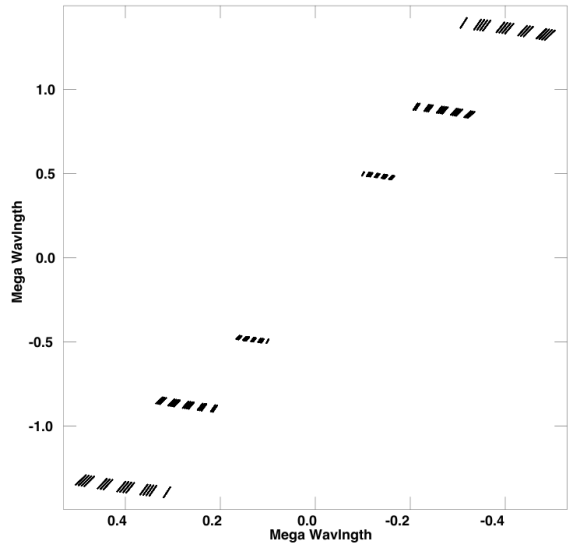}
\caption{Typical uv-coverage of a Cir X-1 e-VLBI session from the campaign in 2009.} 
\label{fig:cirx1_uvplot}
\end{figure}

Fig.~\ref{fig:cirx1_uvplot} shows the typical uv-coverage obtained for a 3-hour observing session. The telescope data streams were correlated with an integration time of 1s and 256 spectral channels across each 16 MHz band. The DiFX correlator output files were analysed with an AIPS-based LBA pipeline written in ParselTongue \citep{Ket06} to obtain preliminary results, following which a more detailed analysis was carried out in AIPS \citep{Gre03} and DIFMAP \citep{she94} using standard data reduction techniques.  Amplitude calibration was carried out in AIPS using the primary flux calibrator J1924-2914 to set the flux density scale. We estimate a systematic uncertainty of about 10 per cent in the overall amplitude scale.

\begin{figure*}
         \centering
          \subfigure [Cir X-1 on 2009 July 15 at 1.7 GHz]{\includegraphics[width=0.3\textwidth]{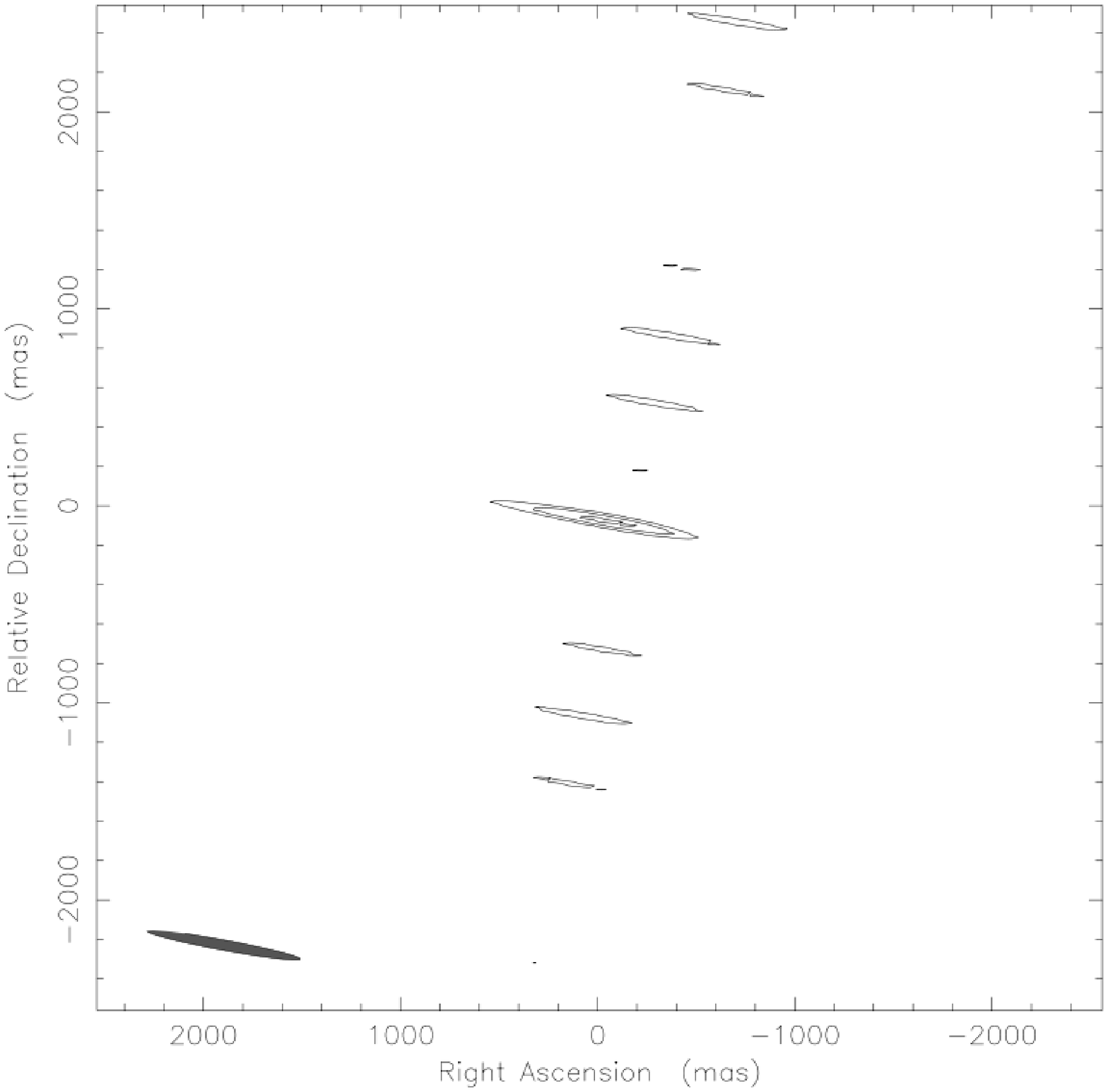}\label{fig:cirx1_map_jul15}}\qquad
         \subfigure [Cir X-1 on 2009 December 9 at 1.4 GHz]{\includegraphics[width=0.3\textwidth]{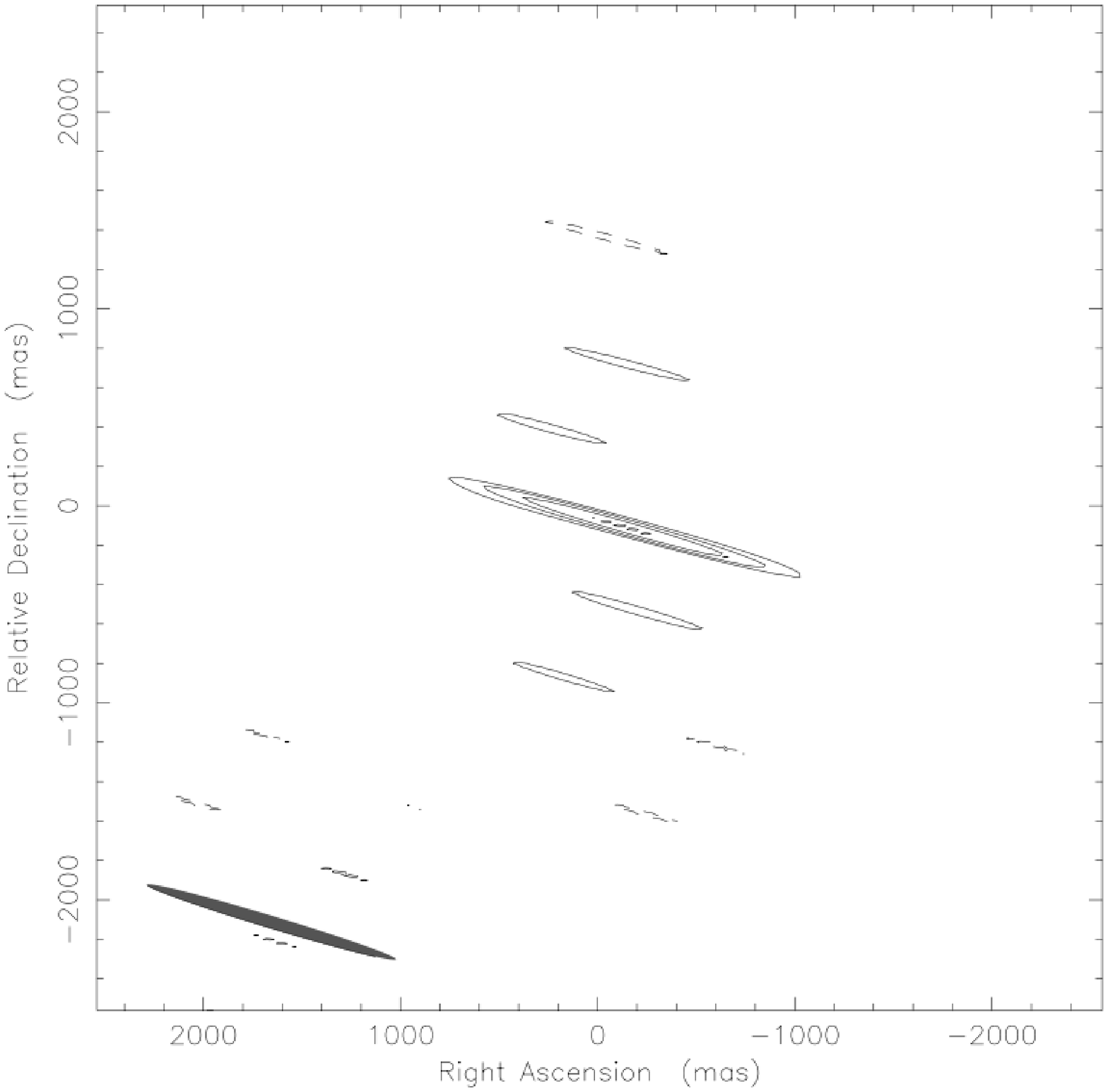}\label{fig:cirx1_map_dec09}}\qquad
         \subfigure [Cir X-1 on 2009 December 11 at 1.4 GHz]{\includegraphics[width=0.3\textwidth]{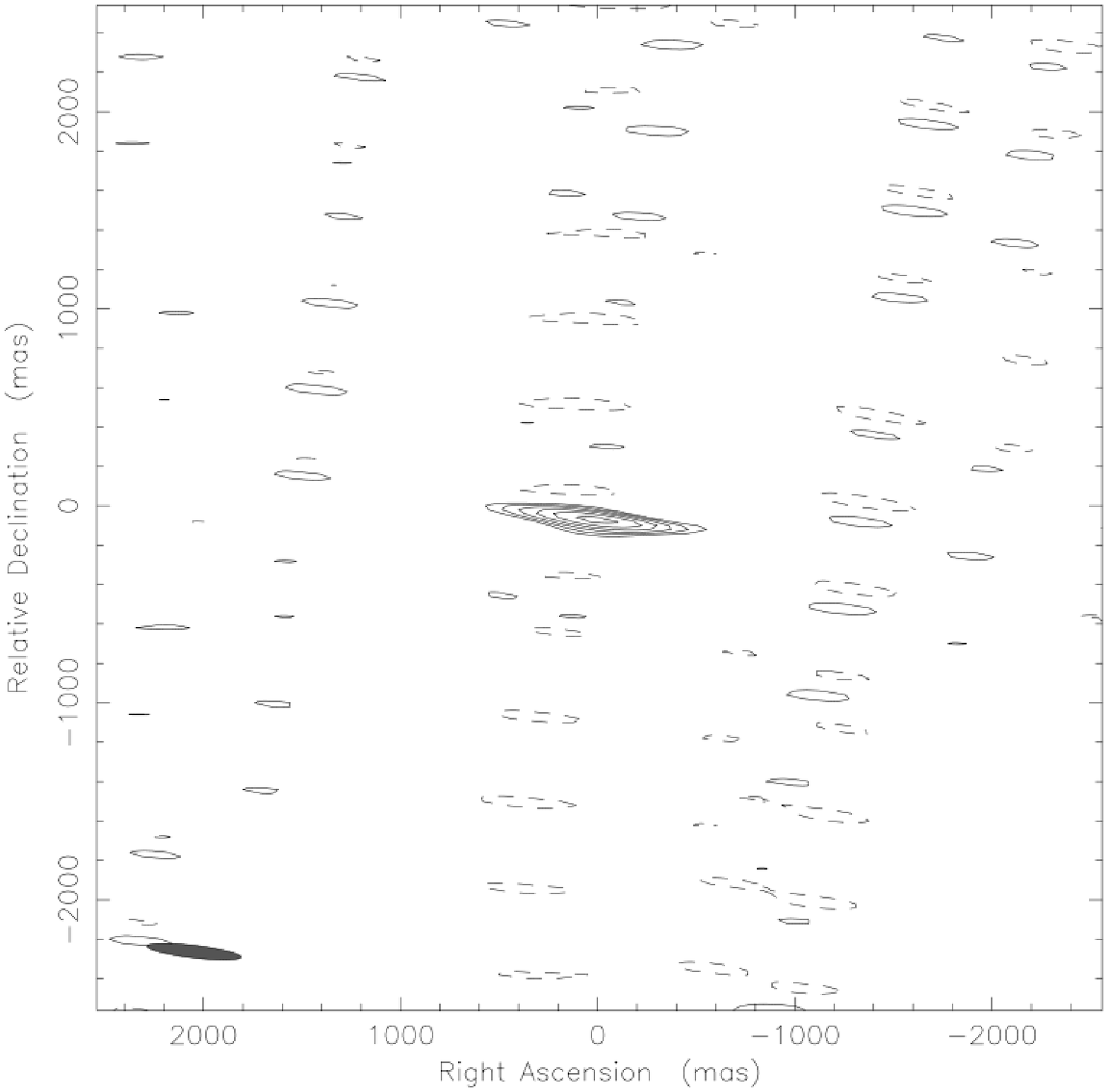}\label{fig:cirx1_map_dec11}}
        
         \caption{Radio images of Cir X-1 during the three epochs when the source was detected in our e-VLBI campaign of 2009.  The source flux densities were 5.6, 3.7 and 2.4\,mJy respectively.  Contours are at levels of -30 and $30(\sqrt{2})^n$\,mJy\,beam$^{-1}$, where $n=0,1,2,...$.  The size of the restoring beam is shown in the bottom left of each panel.}
\end{figure*}

Radio emission associated with Cir X-1 was clearly detected at frequencies of 1.4 and 1.7\,GHz during the observing epochs corresponding to the orbital phases following periastron (orbital phase = 0.0), but no compact radio emission was observed at any other phase along the orbit.  Figures~\ref{fig:cirx1_map_jul15}, \ref{fig:cirx1_map_dec09} and \ref{fig:cirx1_map_dec11} show the LBA images for the three epochs in which the source was detected.  In all cases, the source was unresolved down to the beamsize of the array.  Our best resolution during the epochs when Cir X-1 was detected was $164\times20$\,mas$^2$ in position angle 58.8$^{\circ}$ east of north.  This provides a lower limit on the brightness temperature of $T_{\rm b}>10^6$\,K.  While this does not completely rule out thermal emission, the resolved jets previously detected in this system \citep{Ste93,Hei07} lead us to interpret the emission as synchrotron radiation from the jets.

While simultaneous ATCA data were taken during the e-VLBI observations, the array was in its compact EW352 configuration for all epochs except 2009 July 15, in which it was in an even more compact H75 configuration.  At the low frequency of 1.4\,GHz, the corresponding useful resolutions ($2^{\prime}$ and $7.5^{\prime}$ respectively) would have been insufficient to distinguish emission from the radio jets from persistent diffuse emission from the surrounding radio nebula \citep{Tud06}.  Thus it was not possible to determine whether the long baselines of the LBA were resolving out extended emission on scales of $\sim1^{\prime\prime}$.

\begin{figure*}
  \includegraphics[width=0.5\textwidth,angle=0]{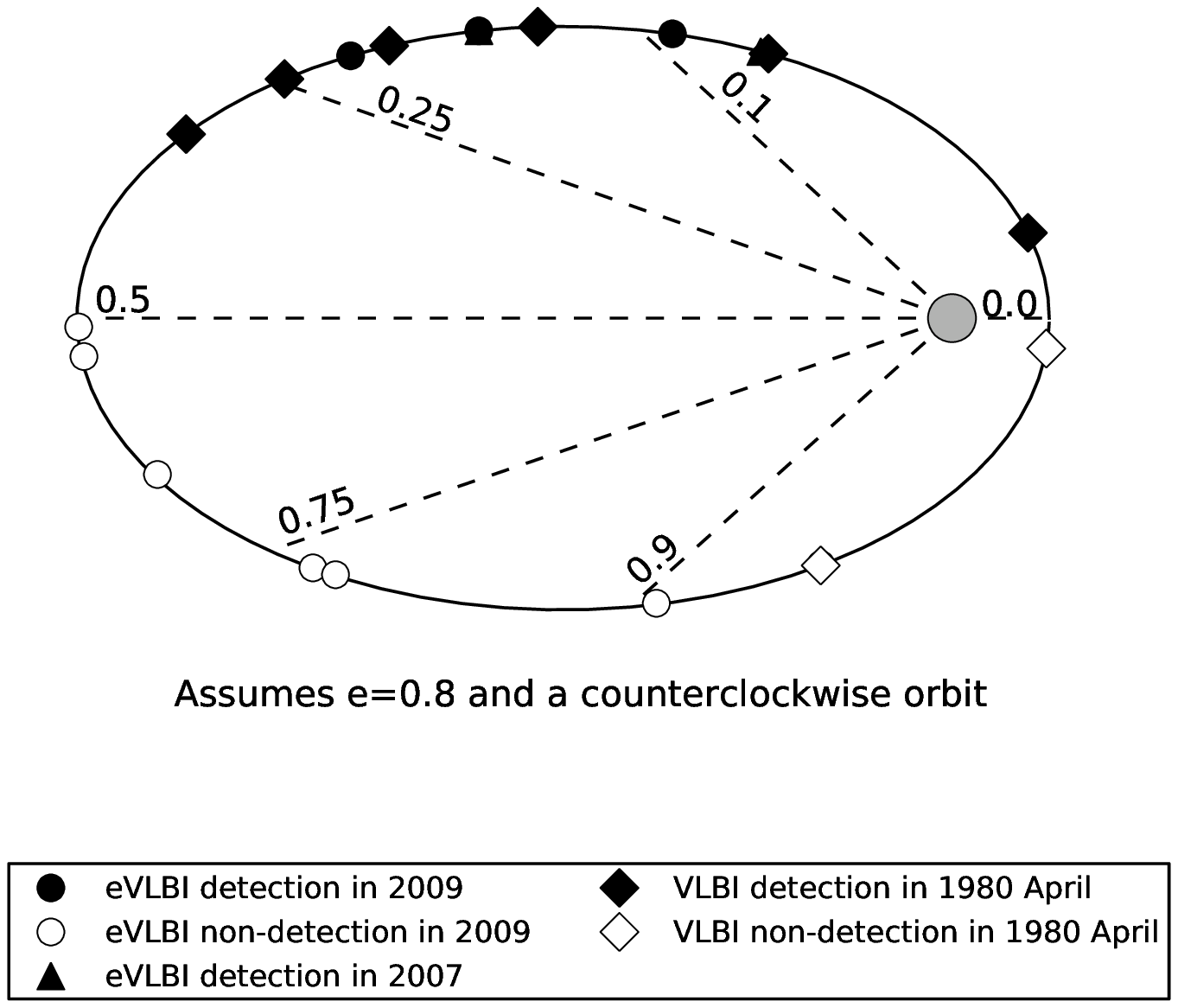}
  \caption{All VLBI observations of Circinus X-1 plotted as a function of position in the orbit.  Filled symbols denote detections and open markers non-detections.  Circles represent the results from our 2009 e-VLBI campaign. Triangles represent the e-VLBI detections of  \citet{Phi07}.  Diamonds represent the VLBI detections made by \citet{Pre83} while Cir X-1 was in a high flaring state.  Dashed lines show the position along the orbit corresponding to the marked orbital phases. There is no evidence for a quiescent component that persists throughout the orbit.}
\label{fig:cirx1_orbit}
\end{figure*}

Fig.~\ref{fig:cirx1_orbit} shows the Cir X-1 orbit and the points along the orbit where the recent e-VLBI observations and the VLBI observations of \citet{Phi07} and \citet{Pre83} were made.  The orbital phase of Circinus X-1 was calculated using the ephemeris of \citet{Nic07}.  The compact radio emission appears to turn on around orbital phase 0.0, but fades away following the flaring event at periastron, and there is no bright, compact radio emission around the rest of the orbit.

To search for radio flaring in the downstream lobes, we made wide-field images of all ten data sets from our e-VLBI campaign, going out to $5^{\prime\prime}$ from the phase centre in each case.  This corresponds to the maximum angular separation between the core and the downstream lobes measured by \citet{Tud08}. By retaining the maximum time resolution and a frequency resolution of 32\,kHz, we were unaffected by either time or bandwidth smearing. No radio emission brighter than $5\sigma$ was detected away from the phase centre (see Table~\ref{tab:evlbi_obs} for the rms noise level in each image). Therefore we rule out any flaring of compact components downstream in the lobes \citep[as seen on larger scales by][]{Fen04} brighter than 2.8 mJy. This implies that either the core flare at periastron did not result in a dark flow propagating downstream at ultrarelativistic speeds to energise the lobes, that the lobe flares are sufficiently faint that they were below our sensitivity limit, or that the lobes are sufficiently diffuse or extended to be resolved out by our LBA observations.

\section{Discussion}

\subsection{The absence of a quiescent compact component}

\begin{figure*}
         \centering
         \subfigure [Radio light curves from our 2009 e-VLBI campaign]{\includegraphics[width=0.45\textwidth]{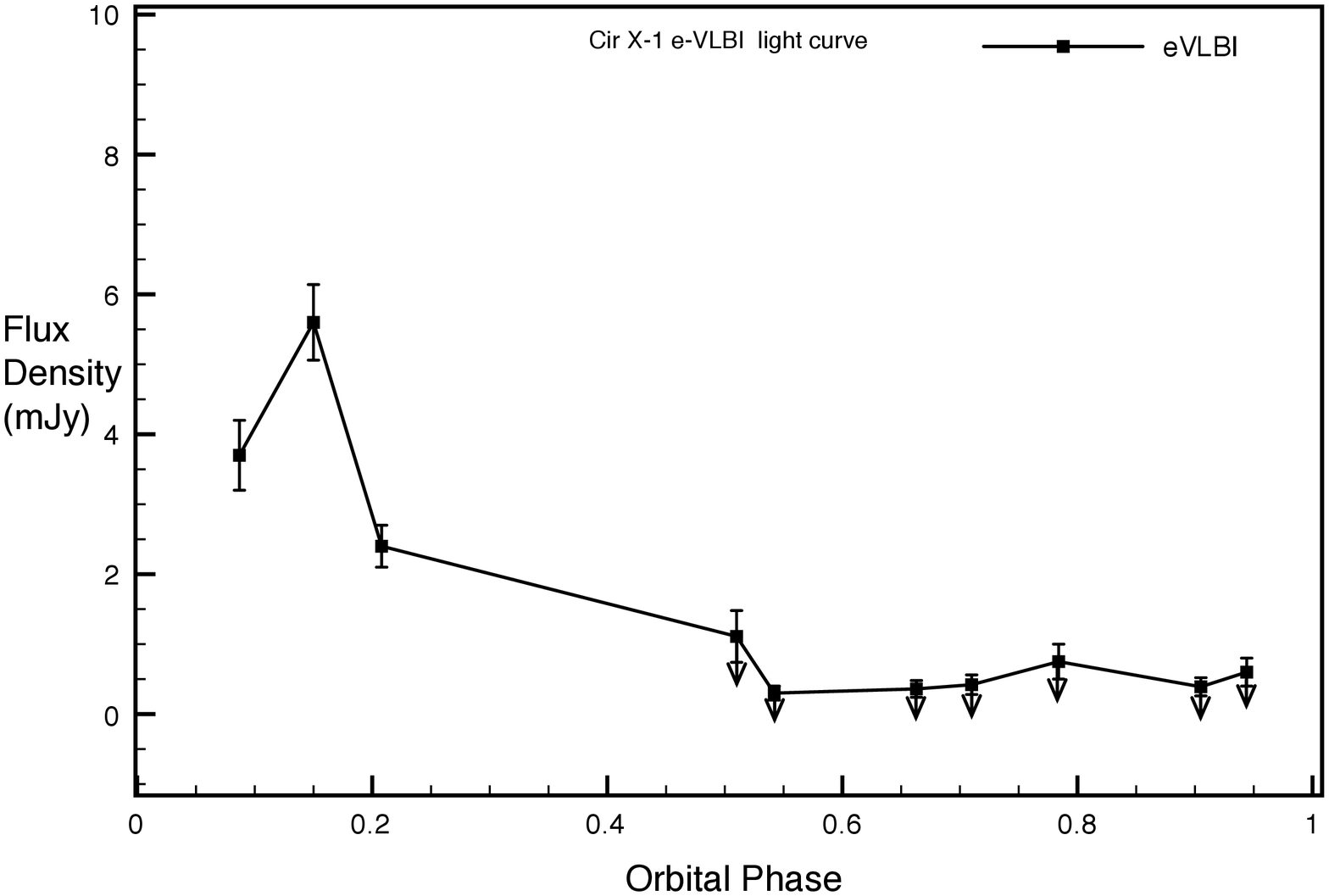}\label{fig:cirx1_lc_10}}\qquad
         \subfigure [Correlated flux density on the Parkes-Tidbinbilla baseline during VLBI observations made in 1980 (data taken from \citet{Pre83}).]{\includegraphics[width=0.45\textwidth]{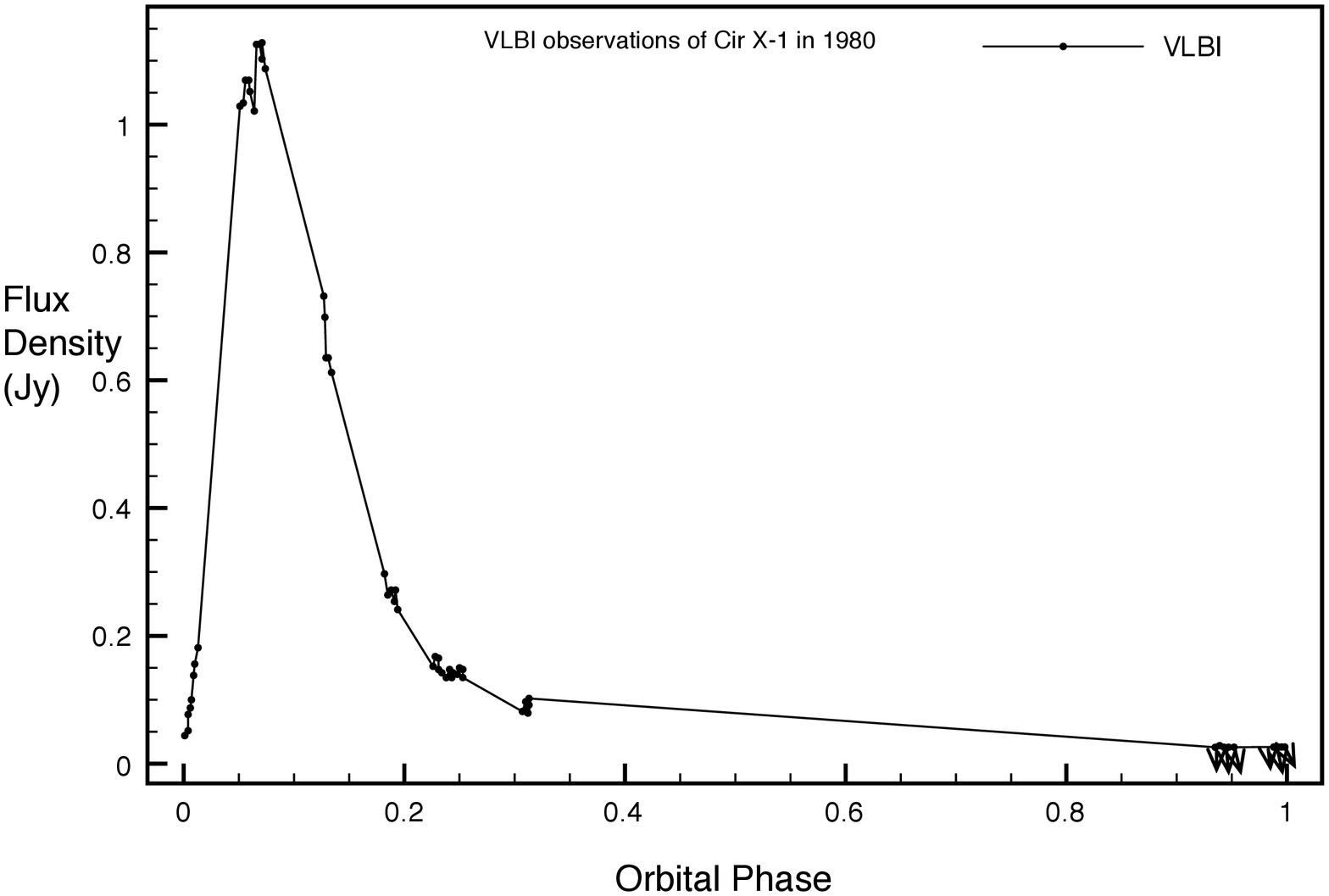}
         \label{fig:cirx1_lc_83}}
         \caption{Radio light curves of Circinus X-1 as a function of orbital phase. Left-hand plot shows data from our e-VLBI campaign, right-hand plot shows the data of \citet{Pre83}.  Note the difference in flux density scales for the two observing campaigns.}
\end{figure*}

Our sampling of the entire orbit of Cir X-1 with high angular resolution e-VLBI observations shows no evidence for compact radio emission outside the flaring event associated with periastron passage (Fig.~\ref{fig:cirx1_lc_10}).  Thus there is no evidence for the compact, non-variable, quiescent component proposed by \citet{Phi07}.  Compact radio emission is detected only at orbital phases following periastron passage when the enhanced mass accretion is believed to power a synchrotron-emitting relativistic jet, as directly resolved on larger scales by \citet{Tud06} and \citet{Tud08}.

Our findings are consistent with all previously-reported VLBI observations of Cir X-1.  \citet{Pre83} observed a large flare in 1980 April (Fig.~\ref{fig:cirx1_lc_83}).  Prior to the beginning of the radio flare at orbital phase 0.002, they did not detect any quiescent emission on the 275-km baseline between Parkes and Tidbinbilla, to a $5\sigma$ upper limit of $\sim20$\,mJy.  All subsequent detections in their monitoring campaign were consistent with the decay of a single large flaring event initiated at or close to orbital phase zero.  The only other reported VLBI detection of Cir X-1, by \citet{Phi07}, was again made in the few days following periastron passage.  Their 1.6-GHz observations were made from orbital phase 0.03--0.11, and their 8.4-GHz observations a day later, from orbital phase 0.14--0.17, in both cases consistent with the results of our observing campaign.

The compact quiescent component postulated by \citet{Phi07} was suggested in response to an apparent discrepancy in the angular size of the source between the observations of \citet{Pre83} in 1980 and those of \citet{Phi07} in 2007.  However, as noted by the authors, the flares being compared were separated by $\sim30$\,y in time and had peak flux densities which differed by almost 2 orders of magnitude.  Therefore, it may not be valid to compare these two very different flares at face value.

The second motivation for the hypothesized compact quiescent component was to explain the absence of resolved jet-like structures in the 1.6-GHz image of \citet{Phi07}.  If the proper motion of the ejecta from the flaring events in Cir X-1 is indeed as high as 400\,mas\,d$^{-1}$, as postulated by \citet{Fen04} and suggested by \citet{Tud08} from a re-analysis of the same data, then any ejecta should have been resolved beyond the beam size in both the observations of \citet{Phi07} and in our e-VLBI runs on 2009 July 15 and December 11, assuming that the proper motion of the ejecta does not vary between outbursts, and that the ejection event occurred at orbital phase 0.0 in each case.  However, we note that this proper motion is based on ATCA observations of correlated radio flaring events seen in the unresolved binary core and downstream jet lobes, and depends on the correct association of a given core flare with the downstream lobe flare.  In no case has a proper motion been definitively measured by fitting the trajectory of moving components.  The most comprehensive analysis of the core-lobe flaring delay was performed by \citet{Tud08}, who considered all possible associations of core and lobe flares at two different frequencies, in three separate outbursts, and concluded a minimum apparent velocity $\beta_{\rm app}c\sim3(d/{\rm kpc})$ where $c$ is the speed of light and $d$ is the source distance, believed to be in the range 3.8--10.5\,kpc \citep{Jon04,Iar05}.  However, we note that since the radio flaring in Circinus X-1 appears to be periodic on the orbital period, this may introduce an ambiguity of some multiple of 16.6\,d into the core-lobe flaring delay, and hence into the assumed proper motion of the ejecta.  Finally, since no moving components were observed in the ATCA images in which the core-lobe delay was detected, the jet responsible for injecting energy into the downstream lobes could be a dark, unseen flow, as suggested to be present in Sco X-1 \citep{Fom01}.  In that case, we would not expect our VLBI observations to resolve discrete ejecta moving from the core to the lobes. The non-detection of lobe emission in any of our VLBI observations suggests that if such a dark flow exists, the working surface where it impacts on the surroundings must be either too diffuse or too faint for us to detect with our observational setup.

In summary, since neither of the original motivations for the existence of a compact, quiescent component can be definitively validated, and in the absence of any observational evidence for non-variable emission of this nature, we can rule out the presence of quiescent radio emission on milliarcsecond scales, to a $3\sigma$ level of 0.3\,mJy\,beam$^{-1}$ in our most sensitive observations (2009 November 30).

\subsection{The absence of secondary flares at apastron}
     
\begin{figure*}
         \centering
         \subfigure [ATCA observations of Circinus X-1 at 4.8 GHz]{\includegraphics[width=0.45\textwidth]{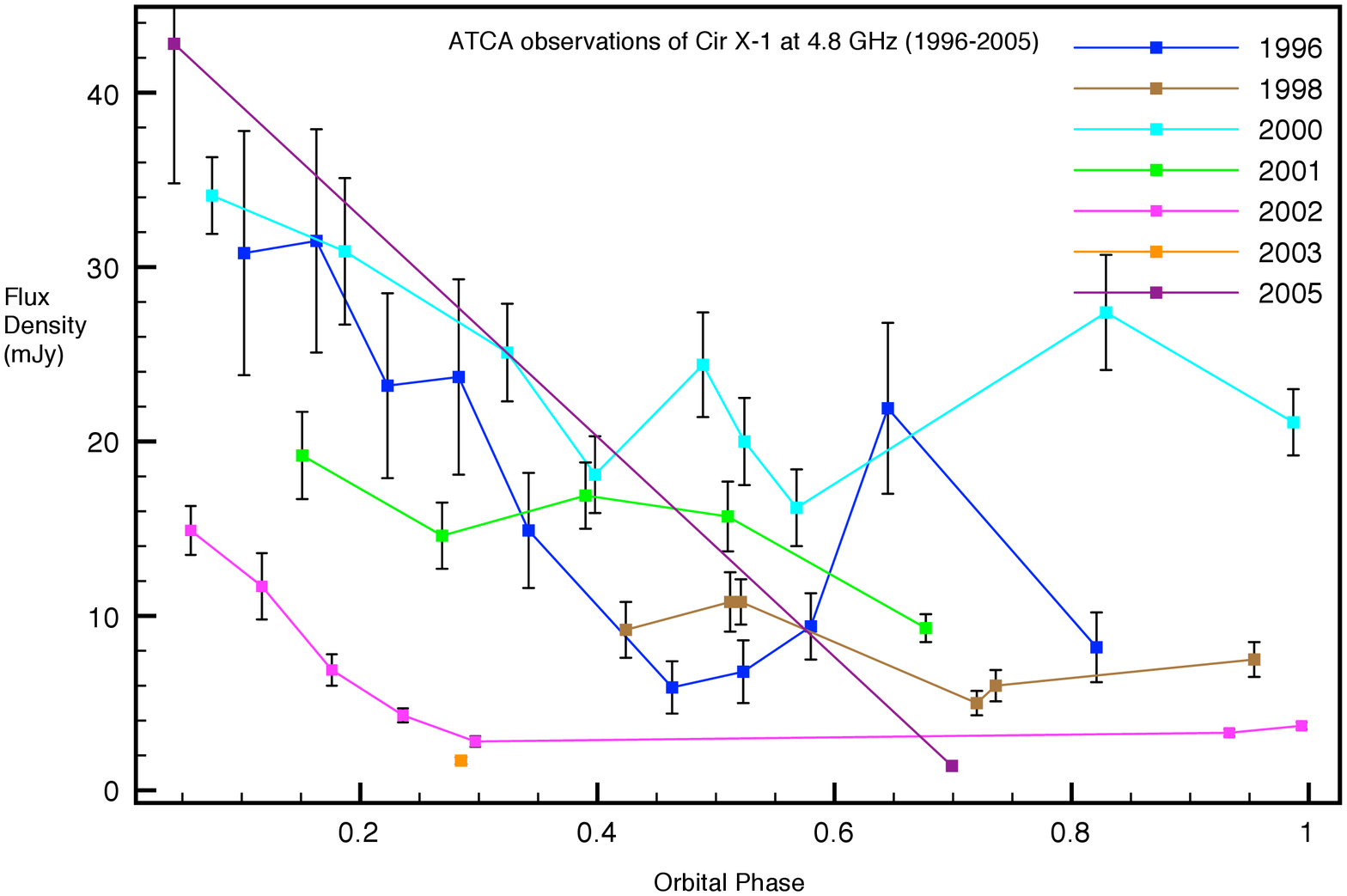}\label{phasing:a}}\qquad
         \subfigure [ATCA observations of Circinus X-1 at 8.6 GHz]{\includegraphics[width=0.45\textwidth]{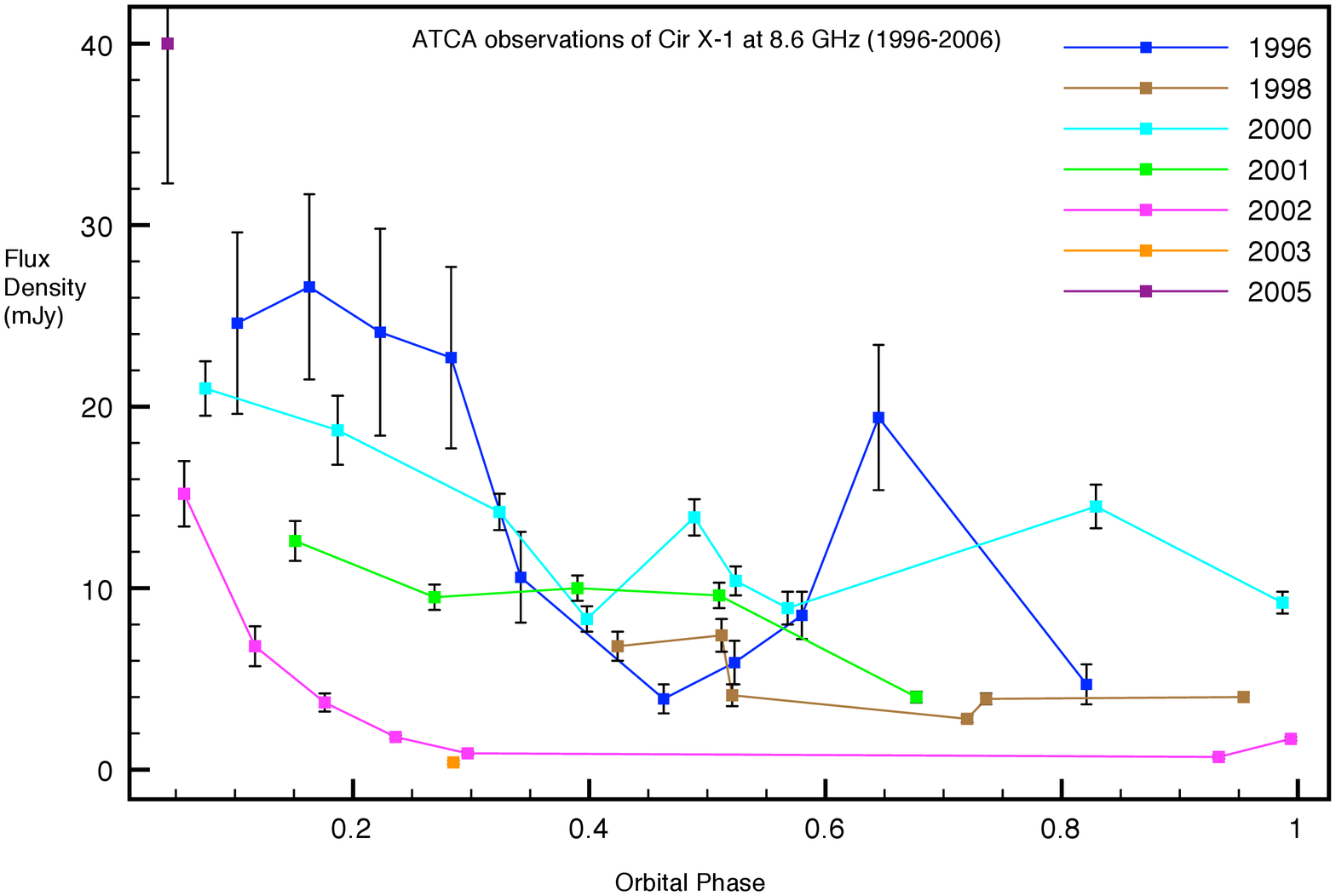}\label{phasing:b}}
         \caption{ATCA observations of Circinus X-1 conducted over a decade (1996-2006), showing radio flux densities as a function of orbital phase \citep[data taken from][]{Tud08}. The left hand plot shows the 4.8-GHz observations and the right hand plot the 8.6-GHz observations.  There is evidence for secondary flares at orbital phase $\sim0.5$.}
\end{figure*}

From the analysis of 10 years' worth of ATCA monitoring data, \citet{Tud08} found evidence for radio flaring events not just close to periastron, but also at phase $0.5\pm0.1$. They attributed this to enhanced wind accretion at a local minimum in the relative velocity between the neutron star and the stellar wind of the companion, close to apastron.  The amplitude of these day-timescale flares was in the range 4--15\,mJy.  Fig.~5 of \citet{Tud08} also shows shorter-timescale flaring events around orbital phase 0.5, lasting only 2--3 hours.  These are most obvious at 8.6\,GHz, being both smoothed out and reduced in amplitude on moving to the lower frequency of 4.8\,GHz.  Figs.~\ref{phasing:a} and \ref{phasing:b} show the daily-averaged data of \citet{Tud08}, with radio flux density plotted as a function of orbital phase, with clear evidence for emission at positions in the orbit away from periastron.  The flares are most evident in the data from 1996 and 2000.  A comparison with the e-VLBI monitoring data in Fig.~\ref{fig:cirx1_lc_10} shows that there is no evidence for similar flaring events at milliarcsecond angular scales in either of the observations within this range of orbital phase (2009 November 30 and December 16).  This implies that these flaring events are either sporadic such that we missed them in two separate orbital cycles, that their peak brightness is sufficiently variable for us not to detect them at our e-VLBI sensitivity limit, or that they arise from sufficiently diffuse emission that they are resolved out on VLBI scales.  The absence of compact emission in our e-VLBI data also rules out the presence of shorter flaring events on timescales of hours, although we note that our observations were carried out at 1.4 and 1.7\,GHz.  Given the smoothing of the short-timescale flares seen with the ATCA at 4.8\,GHz, this is unsurprising.  If these rapid flaring events arise from a compact jet \citep[from which we see emission from the surface of optical depth unity, e.g.][]{Bla79}, it is likely that any rapid variations would be sufficiently smoothed out by the time they reached the 1.4\,GHz photosphere that they would not be significantly detected.

\subsection{The size scale of the emission}

The largest angular scale probed by our VLBI observations corresponds to the length of the shortest baseline (between Mopra and ATCA; 113\,km).  At an observing frequency of 1.4\,GHz, this corresponds to a largest angular size of 412\,mas, or $3300(d/8{\rm kpc})$\,AU, where $d$ is the distance to Cir X-1.  With infinite signal-to-noise, the non-detection of emission away from periastron would therefore limit the size scale of the emission to being greater than this.  If this were solely responsible for the non-detection at orbital phase 0.51 on 2009 December 16, then linear expansion following the detected onset of the flare at orbital phase 0.09 on 2009 December 9 would imply an expansion speed of $>2.7c$.  However, the expansion of the emitting region would reduce the surface brightness, and since the upper limit on the source brightness on 2009 December 16 is only a factor of 2.1 less than the detected source brightness five days earlier, a significantly smaller expansion velocity would be sufficient to render the emission undetectable.  The flux density of a simple adiabatically expanding synchrotron bubble \citep{van66} scales as $R^{-3}$ in the optically thick phase and $R^{-2p}$ in the optically thin phase, where $p$ is the index of the electron energy spectrum, with a canonical value of 2.2.  Thus expansion by only a small factor (1.2--1.3) would be sufficient to cause the observed decrease in source brightness to below the detection limit.  Since the source was unresolved during the flaring event itself, we have no constraints on the original source size, so cannot therefore constrain the expansion velocity.

\section{Conclusions and outlook}

We have used the LBA in e-VLBI mode to monitor the compact radio emission from Circinus X-1 as a function of orbital phase.  The only milliarcsecond-scale detections of the source were made at orbital phases following periastron passage (between phase 0.09 and 0.21).  There is no evidence for compact radio emission during the majority of the orbit, to a $3\sigma$ upper limit of between 0.3 and 1.1\,mJy\,beam$^{-1}$, consistent with previous VLBI observations.  We therefore rule out the hypothesis of \citet{Phi07}, who postulated a non-variable, compact, quiescent component of radio emission with an intrinsic angular size of $\sim35$\,mas.  We find that any flaring events at orbital phase 0.5 must either be too sporadic, faint or diffuse for us to detect them with our e-VLBI observations.  The lack of any resolved jet components moving away from the central binary system suggests that either the ultrarelativistic proper motions reported in the literature are in error, or that the ultrarelativistic flow is dark. If the ultrarelativistic flow exists, the working surface where it impacts the downstream lobes must be either sufficiently faint or diffuse for us not to have detected it at any orbital phase in our e-VLBI observations.

In the near future, higher-frequency 8.4-GHz VLBI observations with a larger array including the ASKAP antenna(s) in Western Australia and the Warkworth Telescope in New Zealand will increase the available resolution by at least a factor of 20. Combined with higher sensitivity, this will enable us to study the morphology of the source and the evolution, dynamics and energetics of a possible relativistic jet associated with Cir X-1 and other such peculiar transient sources.

\section*{Acknowledgments}

The International Centre for Radio Astronomy Research is a Joint Venture between Curtin University of Technology and The University of Western Australia, funded by the State Government of Western Australia and the Joint Venture Partners.  The Australian Long Baseline Array is part of the Australia Telescope which is funded by the Commonwealth of Australia for operation as a National Facility managed by CSIRO.  SJT is a Western Australian Premier's Fellow.  AM is supported by a Curtin University of Technology International Postgraduate Scholarship and is supported as a CSIRO-co-supervised PhD student.

%

\bsp

\label{lastpage}

\end{document}